\documentclass[conference]{IEEEtran}
\IEEEoverridecommandlockouts
\usepackage[utf8]{inputenc}
\usepackage[T1]{fontenc}
\usepackage{cite}
\usepackage{amsmath,amssymb,amsfonts,times}
\usepackage{graphicx}
\usepackage{textcomp}
\usepackage{pgfplots}
\usepackage{scalefnt}
\newcommand{\norm}[1]{\left\lVert#1\right\rVert}

\graphicspath{{figures/}}
\def\BibTeX{{\rm B\kern-.05em{\sc i\kern-.025em b}\kern-.08em
    T\kern-.1667em\lower.7ex\hbox{E}\kern-.125emX}}
\begin{document}

\title{Switched Max-Link Relay Selection Algorithms for Buffer-Aided Relay Systems
}

\author{\IEEEauthorblockN{F. L. Duarte$^{1, 2}$ and R. C. de Lamare$^{1, 3}$}
\IEEEauthorblockA{$^{1}$Centre for Telecommunications Studies (CETUC), Pontifical Catholic University of Rio de Janeiro, Brazil \\
$^{2}$Military Institute of Engineering, IME, Rio de Janeiro, RJ, Brazil  \\
$^{3}$Department of Eletronic Engineering, University of York, United Kingdon \\
Email: \{flaviold, delamare\}@cetuc.puc-rio.br}
}

\maketitle
\linespread{0.93}
\begin{abstract}
In this paper, we investigate relay selection for cooperative
multiple-antenna systems that are equipped with buffers, which
increase the reliability of wireless links. In particular, we
present a novel relay selection technique based on switching and the
selection of the best link, that is named Switched Max-Link. We also introduce
a novel relay selection criterion based on the Maximum Likelihood
(ML) principle and the Pairwise Error Probability (PEP) denoted Maximum Minimum Distance (MMD) that is incorporated
into the proposed Switched Max-Link protocol. We compare the proposed MMD to the existing Quadratic Norm (QN), in terms of  PEP  and computational  complexity. Simulations are then employed to evaluate the performance of
the proposed and existing techniques.  \\
\end{abstract}


\section{INTRODUCTION}

In wireless networks, multipath propagation is a channel propagation
phenomenon that affects the transmission of signals and can be
mitigated through the use of cooperative diversity\cite{f1,f2,f3}.
In cooperative communications with multiple relays, where a number
of relays help a source in transmitting data packets to a
destination, by receiving, processing (decoding) and forwarding
these packets, relay selection schemes are key because of their high
performance \cite{f5, f4, f35}. As cooperative communication can
improve the throughput and extend the coverage of wireless
communications systems, the task of relay selection serves as a
building block to realize it. In this context, relay schemes have
been included in recent/future wireless standards such as Long Term
Evolution (LTE) Advanced \cite{f6, f13} and 5G standards \cite{f12}.

\subsection{Prior and Related Work}

 In conventional relaying, using half duplex (HD) and
decode-and-forward protocols, transmission is usually organized in a
prefixed schedule with two successive time slots. In the first time
slot, the relay receives and decodes the data transmitted from the
source, and in the second time slot the relay forwards the decoded
data to the destination.  Single relay selection schemes use the
same relay for reception and transmission, so they are not able to
simultaneously exploit the best available source-relay (SR) and
relay-destination (RD) channels. The two most common schemes are
bottleneck based and maximum harmonic mean based best relay
selection (BRS) \cite{f5}. The performance of relaying schemes can be improved if
the link with the highest power is used in each time slot.
This can be achieved via a buffer-aided relaying protocol, where
the relay can accumulate packets in its buffer, before
transmitting.
The use of buffers provides an improved
performance and new degrees of freedom for system design \cite{f6, f14}. However,
it suffers from additional delay that must
be well managed for delay-sensitive applications. Buffer-aided
relaying protocols require not only the acquisition of channel state
information (CSI), but control of the buffer status. Some
possible applications of buffer-aided relaying are: vehicular,
cellular, and sensor networks \cite{f6}. In Max-Max Relay Selection (MMRS) \cite{f5}, in the first time slot,
the relay selected for reception can store the received packets in
its buffer and forward them at a later time when selected for
transmission. In the second time slot, the relay selected for
transmission can transmit the first packet in the queue of its
buffer, which was received from the source earlier. MMRS  assumes
infinite buffer sizes. To overcome
this limitation, in  \cite{f5} a hybrid relay selection (HRS) scheme, that is a combination of  BRS and MMRS,
was proposed. Although MMRS and HRS  improve the throughput and/or SNR gain as
compared to BRS, their diversity gain is limited to N (the quantity
of relays). This can be improved by combining adaptive link
selection with MMRS, which results in the
Max-Link \cite{f9} protocol.

The main idea of Max-Link is to select in each
time slot the strongest link among all the available SR and RD links
(i.e., among $2N$ links) for transmission \cite{f7}. For independent and identically distributed (i.i.d.)  links and no
delay constraints, Max-Link achieves a diversity gain of $2N$, which
is twice the diversity gain of BRS and MMRS. Max-Link has been extended in \cite{f11} to account for direct
source-destination (SD) connectivity, which provides resiliency in low transmit SNR conditions \cite{f7}. In \cite{f21,f23,f24,f25}, some buffer-aided relay selection
protocols improve the Max-Link performance by:  reducing the average packet delay, maintaining a good diversity gain, and/or
achieving full diversity gain with a smaller buffer size compared to Max-Link.
In summary, the previous schemes (MMRS, HRS and Max-Link) only use buffer-aided relay
 selection for cooperative single-antenna systems.

\subsection{Contributions}

In this work, we examine buffer-aided relay selection for
cooperative multiple-antenna systems. In particular, we combine the
concept of switching with the concept of selecting the best link used by the Max-Link protocol for cooperative
multiple-antenna systems, which results in the proposed Switched
Max-Link protocol. We also introduce the MMD criterion for selection of relays
 in the proposed scheme, which is based on the ML criterion and the PEP.
The advantage of the MMD algorithm is that it maximizes the minimum value of the PEP argument (PEP worst case).
Simulations illustrate the performance of the proposed  relay
selection techniques. This paper is structured as follows. Section II describes the system
model and the main assumptions made. Section III presents the
proposed Switched Max-Link relay selection protocol and compares the proposed MMD criterion to the existing QN, in terms of PEP and complexity. Section IV illustrates and discusses the simulation results whereas Section V gives the
concluding remarks.

\section{System Description}

\begin{figure}[!h]
\centering
\includegraphics[scale=0.6]{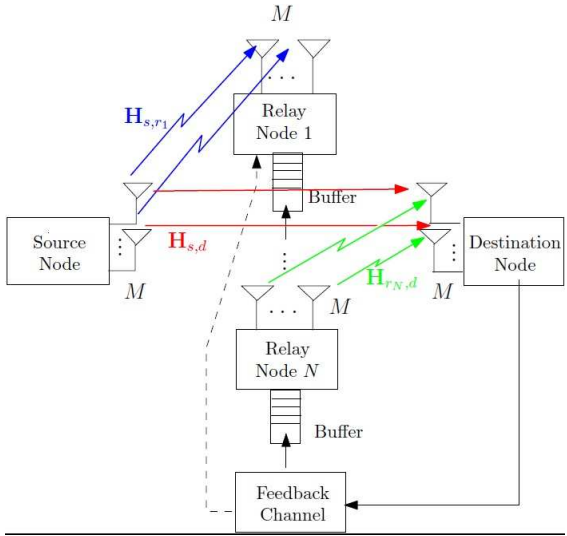}
\caption{System Model}
\label{fig:model}
\end{figure}

We consider a multiple-antenna relay network with one source node, $S$, one
destination node, $ D$, and $N$ half-duplex decode-and-forward (DF)
relays, $R_1$,...,$R_N$. Each relay is equipped with a buffer, whose size is $J$ packets and
each node is equipped with $M$ antennas, and the
transmission is organized in time slots  \cite{f5}. This configuration is considered for simplicity. The considered
system is shown in Fig. \ref{fig:model}.

\subsection{Assumptions}

In cooperative transmissions two time slots are needed to transmit
data packets from the source to the destination, so the energy
transmitted in direct transmission (from the source to
the destination) is twice the energy transmitted in the cooperative
transmission, from the source to the relay selected for reception ($E_s$) or
from the relay selected for transmission to the destination  ($E_{r_j}$),  $E_{r_j}=E_s$. For
this reason, the energy transmitted from each antenna in cooperative transmissions is $E_s/M$ and the energy transmitted from each antenna in direct transmissions is  $2  E_s/M$.

We consider that the channel coefficients are mutually independent
zero mean complex Gaussian random variables (Rayleigh fading).
Moreover, we assume that the transmission is organized in data
packets and the channels are constant for the duration of one packet
and vary independently from one packet to the next. The information about the order of the data packets is contained in
the preamble of each packet, so the original order is restored at
the destination node. Other information such as signaling for
network coordination and pilot symbols for training and knowledge of
the channel (CSI) are also inserted in the preamble of the packet. We consider perfect and imperfect CSI. Furthermore, we assume that the relays do not communicate with each
other.

We also assume that the destination is the central node, being responsible for deciding whether the source or the relay should transmit in a given time slot $i$. The central node has a
perfect channel and buffer state information, so it may run the algorithm in each time slot and select the
relay for transmission or reception through an error-free
feedback channel. This assumption can be ensured by an
appropriate signalling that provides global channel state information
(CSI) at the destination node \cite{f9}. Furthermore, we assume that the source has no CSI and each relay has only information about its SR channel and  buffer status.

\subsection{System Model}

The received signal from the source to the destination  is
organized in an $M \times 1$ vector $\mathbf{y}_{s,d}[i]$ given
by
\begin{eqnarray}
    \mathbf{y}_{s,d}[i]= \sqrt{\frac{2 E_s}{M}} \mathbf{H}_{s,d}\mathbf{x}[i]+\mathbf{n}_d[i],
    \label{eq:1}
\end{eqnarray}
\noindent where $E_s$ represents the total energy of the symbols
transmitted from the source, $\mathbf{x}[i]$ represents
the vector formed by $M$ symbols sent
by the antennas of the source (a symbol of each packet). The quantity
$\mathbf{H}_{s,d}$  represents the $M \times  M$ matrix of $SD$
links and $\mathbf{n}_d$  denotes the zero mean additive white
complex Gaussian noise (AWGN) at the destination receiver.

The received signal from the source to the selected relay is
organized in an $M \times 1$ vector $\mathbf{y}_{s,r_k }[i]$ given
by
\begin{eqnarray}
    \mathbf{y}_{s,r_k}[i]=\sqrt{\frac{E_s}{M}} \mathbf{H}_{s,r_k}\mathbf{x}[i]+\mathbf{n}_{r_k}[i],
    \label{eq:2}
\end{eqnarray}
\noindent where
$r_k$ refers to the selected relay for reception,
$\mathbf{H}_{s,r_k}$ is the $M \times  M$ matrix of $SR_k$ links and
$\mathbf{ n}_{r_k}$ represents the AWGN at the relay selected for
reception.

The signal transmitted from the selected relay and received at the
destination is structured in an $M \times 1$ vector
$\mathbf{y}_{r_j,d }[i]$ given by
    \begin{eqnarray}
    \mathbf{y}_{r_j,d}[i]=\sqrt{\frac{E_{r_j}}{M}}  \mathbf{H}_{r_j,d}\hat{\mathbf{x}}[i]+\mathbf{n}_d[i],
    \label{eq:3}
    \end{eqnarray}
\noindent where $ E_{r_j}$ represents the total energy of the
decoded symbols transmitted from the relay selected for transmission
$r_j$, $\hat{\mathbf{x}}[i]$ is the vector formed by $M$
previously decoded symbols in the relay selected for reception and
stored in its buffer and now transmitted by $r_j$ and $\mathbf{H}_{r_j,d}$ is the $M \times M$ matrix
of  $R_jD$ links.

Assuming perfect CSI, at the relays, we employ the maximum likelihood (ML) receiver \cite{f4}:
    \begin{eqnarray}
    \hat{\mathbf{x}}[i]= \arg \min_{\mathbf{x'}[i]} \left(\norm{\mathbf{y}_{s,r_k}[i]- \sqrt{\frac{E_s}{M}} \mathbf{H}_{s,r_k}\mathbf{x'}[i]}^2\right),
    \label{eq:4}
    \end{eqnarray}
where $\mathbf{x'}[i]$ represents each possible vector
formed by $M$ symbols. As an example, if we have BPSK (number of constellation symbols $N_s= 2$), unit power symbols and $M = 2$, the estimated symbol vector $ \hat{\mathbf{x}}[i]$
may be $[-1~-1]^T$, $[-1~+1]^T$, $[+1~-1]^T$ or $[+1~+1]^T$.

At the destination, we also resort to the ML receiver which
depending on the transmission ($SD$ or $ R_jD$) yields
    \begin{eqnarray}
    \hat{\mathbf{x}}[i]= \arg \min_{\mathbf{x'}[i]} \left(\norm{\mathbf{y}_{s,d}[i]- \sqrt{\frac{2 E_s}{M}} \mathbf{H}_{s,d}\mathbf{x'}[i]}^2\right),
    \label{eq:5}
    \end{eqnarray}
    \begin{eqnarray}
    \hat{\mathbf{x}}[i]= \arg \min_{\mathbf{x'}[i]} \left(\norm{\mathbf{y}_{r_j,d}[i]- \sqrt{ \frac{E_s}{M}} \mathbf{H}_{r_j,d}\mathbf{x'}[i]}^2\right)
    \label{eq:6}
    \end{eqnarray}

The ML receiver of the DF relay looks for an estimate of the vector
of symbols transmitted by the source  $\hat{\mathbf{x}}[i]$,
comparing the quadratic norm between the output $\mathbf{y}_{s,r_k}$
and the term $ \sqrt{E_s/M}\mathbf{H}_{s,r_k}$  multiplied by
$\mathbf{x'}[i]$, that represents each  of the $N_s^M$  possible
transmitted symbols vector $\mathbf{x}$. We compute the symbol
vector which is the optimal solution for the ML rule. The same
reasoning is applied to the ML receiver at the destination. Other
detection techniques can also be employed
\cite{delamare_mber,rontogiannis,delamare_itic,stspadf,choi,stbcccm,FL11,delamarespl07,jidf,jio_mimo,tds,peng_twc,spa,spa2,jio_mimo,P.Li,jingjing,memd,did,bfidd,mbdf,bfidd,mserrr,mmimo,wence,shaowcl08}.

Considering imperfect CSI, a channel error matrix $\mathbf{H}_e$ is added to the channel matrix ($\mathbf{H}_{s,r_k}$,
$\mathbf{H}_{r_j,d}$ or $\mathbf{H}_{s,d}$)  \cite{f16},
where the variance of the $\mathbf{H}_e$ coefficients is given by $\sigma_e^2=\beta E_s^{-\alpha}$ ($\beta \geq 0$ and $0 \leq \alpha \leq 1$), in the case of the channel matrix $\mathbf{H}_{s,r_k}$ or $\mathbf{H}_{r_j,d}$, and $\sigma_e^2=\beta (2 E_s)^{-\alpha}$,  in the case of the channel matrix $\mathbf{H}_{s,d}$.

\section{Proposed Switched Max-Link Relay Selection Protocol}

In this section, we detail the proposed Switched Max-Link relay
selection protocol for cooperative multiple-antenna systems. The
proposed Switched Max-Link  scheme can be implemented by making use
of a network with one source node, $S$,  one destination node, $ D$, and
$N$ half-duplex DF relays, $R_1$,...,$R_N$.  Each relay is equipped with a buffer, whose size is $J$ packets, and
each node is equipped with $M$ antennas, resulting in
a number of $ M  N$ SR channels (links) for
reception, $ M  N$ RD links for
transmission and  $ M$ SD links, as
illustrated in Fig. \ref{fig:model}. This scheme selects the best relay for reception ($R_k$) or the best
relay for transmission ($R_j$) between $N$ relays (the best set of
$M$ $SR$ links among $N$ sets or the best set of  $M$ $RD$ links
among $N$ sets). Similarly to the scheme proposed in \cite{f10}, the MMD relay selection
 criterion (incorporated in Switched Max-Link),  is based on the ML criterion. However, the metrics calculated by MMD are different from those of the scheme in \cite{f10}, which leads to considerably better performance. MMD is also based on the worst case of the  PEP and chooses the relay that has the highest minimum distance. So, it requires calculating the distance between the $N_s^M$
possible vectors of transmitted symbols.

For Switched Max-Link to work properly, it is not necessary that a
certain number of buffer elements be filled with data before the
system starts its normal operation. The buffers may be empty.
Despite of that, in this work, for security, we considered that half
of the buffer elements are filled in an initialization phase
\cite{f5}, by allowing the source to transmit a number of packets to
the relays, before Switched Max-Link is used. During this
initialization phase the relays do not transmit and the source
transmits to the relay with the best set of $M$ SR links among the
available relays.

In each time slot, the proposed Switched Max-Link Protocol may operate in two possible modes ("Direct Transmission" or
"Max-Link"), with three options: a) work in "Direct Transmission" mode, by $S$ sending a
quantity of $M$ packets directly to $D$; b) work in "Max-Link" mode, by $S$ sending a quantity of $M$
packets to $R_k$ and these packets are stored in its buffer; c) work in "Max-Link" mode, by $R_j$
forwarding  a quantity of $M$ packets from its buffer to $D$. Table 1 shows the Switched Max-Link pseudo-code and the following
subsections explain how this protocol works.

\begin{table}[!htb]
\centering
 \caption{Switched Max-Link  Pseudo-Code}
 \label{table1}
\begin{tabular}{l}

\hline
\\

1:   ~~~~ Calculate the metric $\mathcal{D}_{SR_i}$\\

       ~~~~~~      $\mathcal{D}_{SR_i}=  \norm{\sqrt{ E_s/M} \mathbf{H}_{s,r_i}\mathbf{x}_l - \sqrt{ E_s/M} \mathbf{H}_{s,r_i}\mathbf{x}_n}^2$ ; \\
~~~~~~~~~ ~~~~~~~~~~~~~~~~~~~~~~~~~~~~~~~~~~~~~~~~~~~~~~~~~~~~~~~~~ $ i=1,...,N$\\
~~~~~~~~~ ~~~~~~~~~~~~~~~~~~~~~~~~~~~~~~~~~~~~~~~~~~~~~~~~~~~~~~~~~ $ l= 1,...,N_s^M - 1$\\
~~~~~~~~~ ~~~~~~~~~~~~~~~~~~~~~~~~~~~~~~~~~~~~~~~~~~~~~~~~~~~~~~~~~ $ n = l+1,...,N_s^M$ \\

2:  ~~~~  Find the minimum distance - $\mathcal{D}_{\min SR_i}$\\
  ~~~~~~    $ \mathcal{D}_{\min SR_i} = \min{(\mathcal{D}_{SR_i})};$ \\
\\

3:   ~~~~ Calculate the metric $\mathcal{D}_{R_iD}$\\

   ~~~~~~      $\mathcal{D}_{R_iD}=  \norm{\sqrt{ E_s/M} \mathbf{H}_{r_i,d}\mathbf{x}_l - \sqrt{ E_s/M} \mathbf{H}_{r_i,d}\mathbf{x}_n}^2$ ; \\
~~~~~~~~~ ~~~~~~~~~~~~~~~~~~~~~~~~~~~~~~~~~~~~~~~~~~~~~~~~~~~~~~~~~ $ i=1,...,N$\\
~~~~~~~~~ ~~~~~~~~~~~~~~~~~~~~~~~~~~~~~~~~~~~~~~~~~~~~~~~~~~~~~~~~~ $ l= 1,...,N_s^M - 1$\\
~~~~~~~~~ ~~~~~~~~~~~~~~~~~~~~~~~~~~~~~~~~~~~~~~~~~~~~~~~~~~~~~~~~~ $ n = l+1,...,N_s^M$ \\

4:  ~~~~  Find the minimum distance - $\mathcal{D}_{min R_iD}$\\
  ~~~~~~    $ \mathcal{D}_{\min R_iD} = \min{(\mathcal{D}_{R_iD})};$ \\
\\
5:  ~~~~  Perform ordering on $\mathcal{D}_{\min SR_i}$ and $\mathcal{D}_{\min R_iD}$\\
\\
6:  ~~~~  Find the maximum minimum distance (considering the buffer status)\\
~~~~~~ $\mathcal{D} _{\max \min SR-RD}= \max{(\mathcal{D}_{\min SR_i},\mathcal{D}_{\min R_iD})}$;
\\
\\

7:   ~~~~ Calculate the metric $\mathcal{D}_{SD}$\\

       ~~~~~~      $\mathcal{D}_{SD}=  \norm{\sqrt{2 E_s/M} \mathbf{H}_{s,d}\mathbf{x}_l - \sqrt{ 2 E_s/M} \mathbf{H}_{s,d}\mathbf{x}_n}^2;$ \\
~~~~~~~~~ ~~~~~~~~~~~~~~~~~~~~~~~~~~~~~~~~~~~~~~~~~~~~~~~~~~~~~~~~~ $ l= 1,...,N_s^M - 1$\\
~~~~~~~~~ ~~~~~~~~~~~~~~~~~~~~~~~~~~~~~~~~~~~~~~~~~~~~~~~~~~~~~~~~~ $ n = l+1,...,N_s^M$ \\

8:  ~~~~  Find the minimum distance - $\mathcal{D}_{\min SD}$\\
 ~~~~~~    $ \mathcal{D}_{\min SD} = \min{(\mathcal{D}_{SD})};$ \\

\\

9:  ~~~~  Select the transmission mode\\

 ~~~~~~   $\mathbf{If}$  $\mathcal{D}_{\min SD} \geq \mathcal{D}_{\max \min SR-RD}$\\

 ~~~~~~   ~~    Operate in "Direct transmission mode"; \\

 ~~~~~~     ~~  $\mathbf{ else}$ \\

 ~~~~~~   ~~~~Operate in "Max-Link mode" ;\\

 ~~~~~~   $\mathbf{end}$

\end{tabular}
\end{table}

\subsection{Calculation of relay selection metric}

In the first step we calculate the metric $\mathcal{D}_{SR_i}$
related to the $SR$ channels of each relay $R_i$ in Max-Link mode:
\begin{eqnarray}
  \mathcal{D}_{SR_i}=  \norm{\sqrt{ \frac{E_s}{M}} \mathbf{H}_{s,r_i}\mathbf{x}_l - \sqrt{ \frac{E_s}{M}} \mathbf{H}_{s,r_i}\mathbf{x}_n}^2,
  \label{eq:7}
\end{eqnarray}
where "$l$" is different from "$n$",  $\mathbf{x}_l$  and $\mathbf{x}_n$
represent each possible vector formed by $M$ symbols.

This metric is calculated for each of the $C_2^{N_s^M}$
(combination of $N_s^M$  in $2$) possibilities. As an example, if
$M = 2$ and $N_s=2$, we have $C_2^4= 6$ possibilities. Then, we store the information
related to the smallest metric ($\mathcal{D}_{min SR_i}$), for being
critical (a bottleneck) in terms of performance, and thus each relay
will have a minimum distance associated with its SR channels.

In the second step we calculate the metric $\mathcal{D}_{R_iD}$
related to the $RD$ channels of each relay $R_i$:
\begin{eqnarray}
    \mathcal{D}_{R_iD}=  \norm{\sqrt{ \frac{E_s}{M}} \mathbf{H}_{r_i,d}\mathbf{x}_l - \sqrt{\frac{ E_s}{M}} \mathbf{H}_{r_i,d}\mathbf{x}_n}^2,
    \label{eq:8}
\end{eqnarray}
where "$l$" is different from "$n$". This metric is calculated for each
one of the $C_2^{N_s^M}$  possibilities. Then, we store the information related
to the minimum distance ($\mathcal{D}_{\min R_iD}$), and thus each
relay will have a minimum distance associated with its RD channels.

In the third step, after calculating the metrics $\mathcal{D}_{\min
SR_i}$  and $\mathcal{D}_{ \min R_iD}$ for each of the relays, as
described previously, we look for the largest value of the
minimum distance:
\begin{eqnarray}
    \mathcal{D} _{\max \min SR-RD}= \max(\mathcal{D}_{\min SR_i}, \mathcal{D}_{\min R_iD}),
    \label{eq:9}
    \end{eqnarray}
where "$i$" is the index of each relay ($1,2,...,N$). Therefore, we
select the relay that is associated with this $\mathcal{D} _{\max\min
SR-RD}$, considering its buffer status. This relay will be selected for reception (if its buffer is not full) or transmission (if its buffer is not empty),
depending on this metric  is associated with the SR or RD channels,
respectively.

\subsection{Calculation of the metric for direct transmission}

In this step we calculate the metric $\mathcal{D}_{SD}$ related to
the $SD$ channels for the direct transmission mode:
\begin{eqnarray}
    \mathcal{D}_{SD}=  \norm{\sqrt{\frac{2 E_s}{M}} \mathbf{H}_{s,d}\mathbf{x}_l - \sqrt{\frac{2 E_s}{M}} \mathbf{H}_{s,d}\mathbf{x}_n}^2,
    \label{eq:10}
\end{eqnarray}
where "$l$" is different from "$n$". This metric is calculated for each
of the $C_2^{N_s^M}$  possibilities. Then, we store the information related to the
minimum distance ($\mathcal{D}_{\min SD}$),  associated with SD
channels.

\subsection{Comparison of metrics and choice of transmission mode}

After calculating all the metrics associated to the SR and RD
channels, finding $\mathcal{D}_{\max \min SR-RD}$ and calculating the
metrics associated to the SD channels and finding $\mathcal{D}_{\min
SD}$, we compare these parameters and select the transmission mode:

    - If $\mathcal{D}_{\min SD} \geq \mathcal{D}_{\max\min SR-RD}$, we select "Direct transmission mode",

    - Otherwise, we select "Max-Link mode".

If we do not consider the possibility of a direct SD connectivity ("Direct
Transmission mode"), considering only the cooperative SR-RD connectivity ("Max-Link mode"), we have another scheme,
called "MMD-Max-Link",  instead of the proposed "Switched Max-Link" scheme.

\subsection{Pairwise Error Probability}

We can simplify the equations in (\ref{eq:7}),  (\ref{eq:8}) and (\ref{eq:10}), making $\mathcal{D}=  E_s/M \times \mathcal{D'}$, where $\mathcal{D'}$=$\norm{\mathbf{H}(\mathbf{x}_n-\mathbf{x}_l)}^2$,  for $\mathcal{D}_{SR_i}$ and $\mathcal{D}_{R_iD}$, or $\mathcal{D'}$=~2$\norm{\mathbf{H}(\mathbf{x}_n-\mathbf{x}_l)}^2$ , for $\mathcal{D}_{SD}$. We know that the PEP considers the error event when $\mathbf{x}_n$  is transmitted and the detector computes an incorrect $\mathbf{x}_l$  (where "$l$" is different from "$n$"), based on the received symbol \cite{f26}. The PEP is given by

\begin{eqnarray}
\begin{split}
\mathbf{P}(\mathbf{x}_n \rightarrow \mathbf{x}_l | \mathbf{H})&= Q\left(\sqrt{\frac{E_s}{2 N_0M} \mathcal{D'}}\right)
  \label{eq:19}
\end{split}
\end{eqnarray}
where $N_0$ is the AWGN noise spectrum density.

We may consider that the PEP will have its maximum value for the minimum value of $\mathcal{D'}$ (worst case of the PEP).
So, for the worst case of the PEP ($\mathcal{D'}_{\min}$), in direct SD transmissions, in each time slot, we have

\begin{eqnarray}
\mathbf{P}(\mathbf{x}_n \rightarrow \mathbf{x}_l | \mathbf{H})= Q\left(\sqrt{\frac{E_s}{2 N_0M} \mathcal{D'}_{\min}}\right)
  \label{eq:101}
\end{eqnarray}

However, for cooperative SR-RD transmissions, an approximated expression for computing the worst case of the PEP in each time slot (regardless of whether it is an SR or RD link) is given by
\begin{eqnarray}
\mathbf{P}(\mathbf{x}_n \rightarrow \mathbf{x}_l | \mathbf{H})\approx 1- \left(1-Q\left(\sqrt{\frac{E_s}{2 N_0M} \mathcal{D'}_{\min}}\right)\right)^2
  \label{eq:102}
\end{eqnarray}

The advantage of the MMD algorithm compared to QN is that  MMD maximizes the metric $\mathcal{D'}_{\min}$  and QN does not take it into account. The QN algorithm is  based only on the total power of these links (as the traditional Max-Link). Its metric $\mathcal{Q}$ is related to the quadratic norm of each matrix $\mathbf{H}$, and the matrix selected by this criterion is: $\mathbf{H}^{QN}= \arg \max_{\mathbf{H}} \norm{\mathbf{H}}^2$. Even though the QN criterion selects the relay that
has the largest quadratic norm of the channel coefficients matrices, the minimum value of the PEP argument $\mathcal{D'}_{\min}^{QN}$ associated with $\mathbf{H}^{QN}$, selected by the QN criterion,  may be not as high as the minimum value of the PEP argument $\mathcal{D'}_{\min}^{MMD}$ associated with $\mathbf{H}^{MMD}$, selected by the MMD criterion. So, we have

\begin{eqnarray}
\mathbf{P}^{MMD}(\mathbf{x}_n \rightarrow \mathbf{x}_l | \mathbf{H}^{MMD}) \leq \mathbf{P}^{QN}(\mathbf{x}_n \rightarrow \mathbf{x}_l | \mathbf{H}^{QN})
  \label{eq:22}
\end{eqnarray}
where $\mathbf{P}^{MMD}(\mathbf{x}_n \rightarrow \mathbf{x}_l | \mathbf{H}^{MMD})$ is the PEP for the worst case in the MMD criterion and $\mathbf{P}^{QN}(\mathbf{x}_n \rightarrow \mathbf{x}_l | \mathbf{H}^{QN})$ is the PEP for the worst case in the QN criterion.

\subsection{Complexity}

As we have seen, the metric $\mathcal{D}$ may be calculated for each of the $C_2^{N_s^M}$  possibilities. However, it is not necessary to calculate all of them. We may generalize the total number  $\mathcal{X}$  of calculations of the  metric $\mathcal{D}$,  needed by the MMD criterion, for each matrix $ \mathbf{H}$:

\begin{eqnarray}
\begin{split}
\mathcal{X}&=WC_1^{M}+2 W^2 C_2^{M}+4 W^3  C_3^{M}+...+2^{M-1}W^M C_M^{M}\\
&=  \sum_{i=1}^{M} 2^{i-1}  W^i C_i^{M}
\label{eq:23}
\end{split}
\end{eqnarray}
where $W$ is the total number of different distances between the constellation symbols. If we have BPSK, $W=1$, and QPSK, $W=3$.

\begin{table}[!htb]
\centering
 \caption{Maximum Minimum Distance versus Quadratic Norm - Complexity}
 \label{table2}
\begin{tabular}{l|ll}
\hline
Operations/Criterion& Maximum Minimum Distance & Quadratic Norm\\
\hline
additions & $2N M(\mathcal{X}-1)$ & $2N (M^2-1)$\\
\hline
 multiplications &  $2NM\mathcal{X}$ & $2N M^2$\\
\hline

\end{tabular}
\end{table}

\begin{figure}[!htb]
\centering
\includegraphics[scale=0.5]{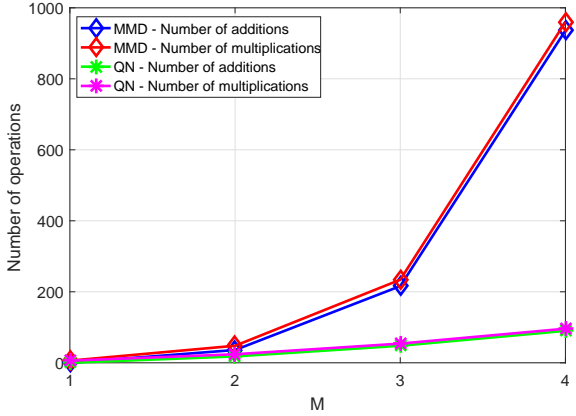}
\caption{MMD-Max-Link and QN-Max-Link complexity.}
\label{fig:complexityMMDMaxLink}
\end{figure}

Table \ref{table2} shows the complexity of the MMD and QN criteria for a number of $N$ relays, $M$ antennas, considering only the cooperative transmission (not considering the direct transmission mode), and the constellation type.

Fig. \ref{fig:complexityMMDMaxLink} shows the complexity of the MMD and QN criteria, for $N=3$ (a source, 3 relays and a destination), and BPSK.  By the analysis of this result,  it is observed that the complexity of the MMD criterion with $M=2$  is not so higher than the complexity of the QN criterion. If we increase the number of antennas to $M=3$ (or more)  in each node, the complexity of MMD criterion becomes considerably higher than the complexity of QN criterion.

\section{Simulation Results}

This section illustrates and discusses the
simulation results of the  proposed "Switched Max-Link", the
"MMD-Max-Link",  the "conventional MIMO" (direct transmission, without relaying) and the Max-Link with the QN criterion ("QN-Max-Link").  QN-Max-Link with a single antenna is equal to the traditional Max-Link. We assume that the transmitted signals belong to BPSK or QPSK
constellations. The 16-QAM constellation was not included in this work because of its higher complexity.  Each relay is equipped with a buffer whose size is $J=4 $ packets. Note that we tested the performance for different $J$ but found that $J=4$  is sufficient to ensure a good performance.
We also assume unit power channels ($\sigma_{ s,r}^2$
$=$ $\sigma_{ r,d}^2$ $=$  $\sigma_{ s,d}^2$ $= 1$), $N_0 =1$ and $E_S = E_{r_j} = E$ (total energy transmitted). The transmit signal-to-noise ratio SNR ($E/N_0$) ranges
from 0 to 12 dB and the performances of the transmission schemes
were tested for 20000 packets, each containing 100 symbols.

\begin{figure}[!h]
\centering
\includegraphics[scale=0.5]{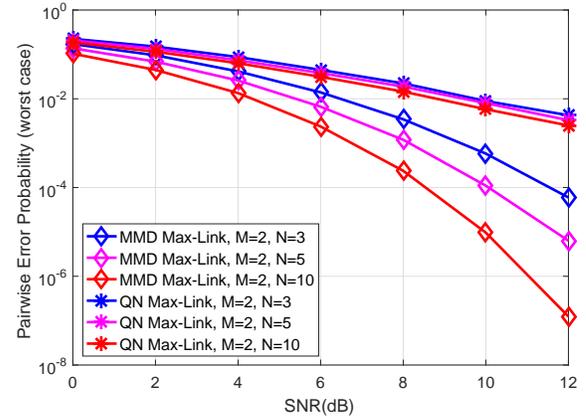}
\caption{MMD-Max-Link and QN-Max-Link PEP  performance.}
\label{fig:pepMMDMaxLink}
\end{figure}

Fig. \ref{fig:pepMMDMaxLink} shows the PEP performance of the MMD-Max-Link and QN-Max-Link protocols, for $M = 2$, $N$ = 3, 5 and 10, BPSK and perfect CSI.  By the analysis of this result,  it is observed, as expected, that  for multiple antennas the performance of the MMD-Max-Link scheme is much better than the performance of QN-Max-Link  for the total range of SNR tested. When we increase $N$, the MMD-Max-Link has its performance improved. The same does not happen to QN-Max-Link,  as the QN criterion does not take the metric $\mathcal{D'}_{\min}$  into account.

\begin{figure}[!h]
\centering
\includegraphics[scale=0.5]{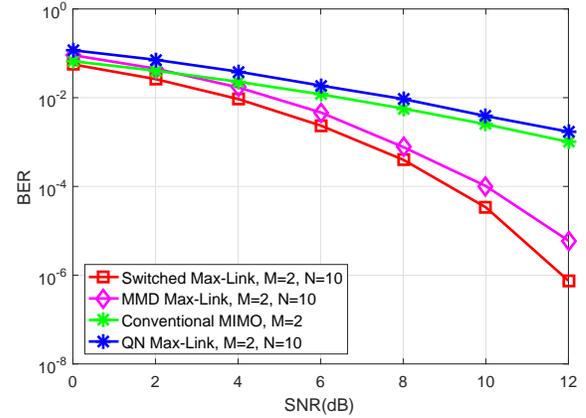}
\caption{Switched Max-Link, MMD-Max-Link, Conventional
MIMO (direct transmission) and QN-Max-Link BER  performance.}
\label{fig:berMaxlink}
\end{figure}

Fig. \ref{fig:berMaxlink} shows the Switched Max-Link, the MMD-Max-Link, the QN-Max-Link
and the conventional MIMO (direct transmission) BER  performance
comparison for $M= 2$,  $N= 10$, BPSK and perfect CSI. We notice that the
performance of the MMD-Max-Link scheme  is worse than
the performance of the conventional MIMO scheme for a SNR less than
2 dB. Nevertheless, the performance of the proposed Switched
Max-Link scheme is better than the performance of the
conventional MIMO for a wide range of SNR values. It is observed, as expected, that the performance of the
proposed Switched Max-Link scheme is better than the performance
of the MMD-Max-Link scheme, as well as its resiliency in
low transmit SNR conditions.

\begin{figure}[!h]
\centering
\includegraphics[scale=0.5]{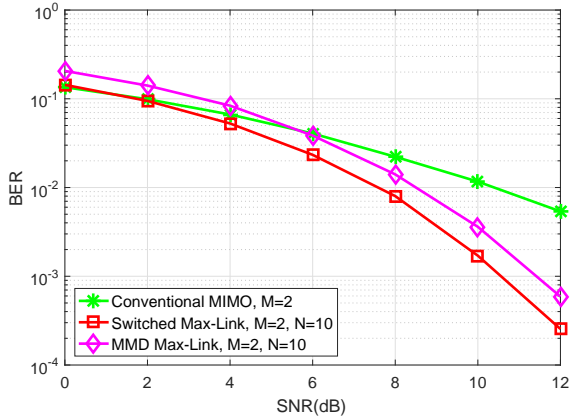}
\caption{Switched Max-Link, MMD-Max-Link and Conventional
MIMO (direct transmission)  BER  performance.}
\label{fig:serMaxlink}
\end{figure}

Fig. \ref{fig:serMaxlink} shows the Switched Max-Link, the MMD-Max-Link and the conventional MIMO BER performance comparison for $M= 2$,  $N= 10$, QPSK and perfect CSI (the QN-Max-Link was not considered as its performance is worse than the performance of the proposed protocol).
The performance of the MMD-Max-Link scheme is worse than
the performance of the conventional MIMO scheme for a SNR less than
6 dB. Nevertheless, the performance of the proposed Switched
Max-Link scheme is better than the performance of the
conventional MIMO for a wide range of SNR values. It is observed  that the performance of the
proposed Switched Max-Link scheme is better than the performance
of the MMD-Max-Link scheme, as well as its resiliency in
low transmit SNR conditions.

\begin{figure}[!h]
\centering
\includegraphics[scale=0.5]{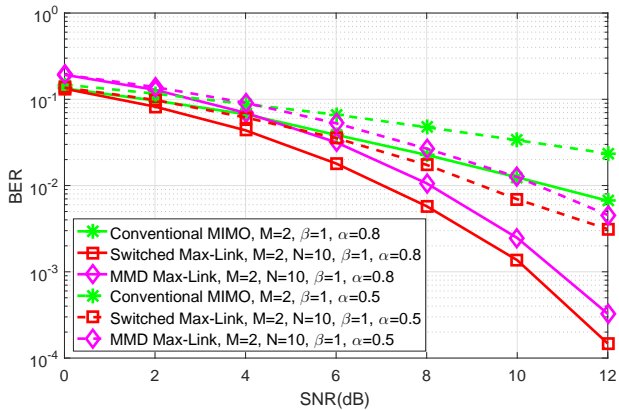}
\caption{Switched Max-Link, MMD-Max-Link and Conventional
MIMO (direct transmission) BER  performance for imperfect channel knowledge.}
\label{fig:berMaxlinkerro}
\end{figure}

Fig. \ref{fig:berMaxlinkerro} shows the Switched Max-Link, the MMD-Max-Link
and the conventional MIMO  BER  performance
comparison for $M= 2$,  $N= 10$, BPSK and  imperfect channel knowledge ($\beta=1$, $\alpha=0.5$ and $\alpha=0.8$). As in the case of perfect channel knowledge, the
performance of the  proposed Switched
Max-Link scheme is better than the performance of the
conventional MIMO for a wide range of SNR values. It is observed  that the performance of the
proposed Switched Max-Link scheme is still better than the performance
of the MMD-Max-Link scheme, as well as its resiliency in
low transmit SNR conditions.

\section{Conclusions}

In this paper we have presented the benefits of using
buffers and multiple antennas for the design of half-duplex
decode-and-forward relaying protocols in cooperative communication
systems, by using the MMD relay selection criterion, based on the ML criterion and the PEP. Moreover, a new
cooperative  protocol using multiple antennas that combines the concept of switching
and  the concept of selection of the best link used by Max-Link and incorporates the MMD selection criterion has been proposed. The proposed Switched Max-Link was evaluated
experimentally and outperformed the conventional direct transmission
and the existing QN Max-Link scheme.

\end{document}